\documentstyle[11pt]{article}
\newcommand{\vol}{\int \mbox{d} ^ n x}
\newcommand{\beqn}{\begin{equation}}
\newcommand{\eeqn}{\end{equation}}
\newcommand{\barr}{\begin{eqnarray}}
\newcommand{\earr}{\end{eqnarray}}
\newcommand{\e}{\varepsilon}
\newcommand{\p}{\partial}
\newcommand{\nn}{\nonumber}
\def\s0#1#2{\mbox{\small{$\frac{#1}{#2}$}}}
\def\f#1#2#3{\lambda_{#1#2}^{#3}}
\def\GP{\epsilon}
\def\platz{b}
\begin{document}
\begin{titlepage}

\begin{flushright}

KUL--TF--96/15\\
ULB--TH--96/10\\
hep-th/9606172\\
June 1996\\

\end{flushright}
\vfill

\begin{center}
{\Large{\bf  Global Symmetries in the Antifield--Formalism}}
\end{center}
\vfill

 \begin{center}
 {\large
 Friedemann Brandt\,$^{a,1}$,
 Marc Henneaux\,$^{b,c,2}$\ and
 Andr\'{e} Wilch\,$^{b,3}$}
 \end{center}
 \vfill

 \begin{center}{\sl
 $^a$ Instituut voor Theoretische Fysica, Katholieke
 Universiteit Leuven,\\
 Celestijnenlaan 200 D, B--3001 Leuven, Belgium\\[1.5ex]
 
 $^b$ Facult\'e des Sciences, Universit\'e Libre de Bruxelles,\\
 Campus Plaine C.P. 231, B--1050 Bruxelles, Belgium\\[1.5ex]

 $^c$ Centro de Estudios Cient\'\i ficos de Santiago,\\
 Casilla 16443, Santiago 9, Chile

 }\end{center}
 \vfill

\begin{abstract}
In this paper, two things are done. (i) First, it is shown that
any global symmetry
of a gauge-invariant theory can be extended to the ghosts and 
the antifields so as
to leave invariant the solution of the master-equation (before
gauge fixing).
(ii) Second, it is proved
that the incorporation of the rigid symmetries to 
the solution of the master-equation
through the introduction of a constant ghost
for each global symmetry can be
obstructed already at the classical
level whenever the theory possesses higher order 
conservation laws. Explicit examples are given.
\end{abstract}

\vspace{5em}

\hrule width 5.cm
\vspace*{.5em}

{\small \noindent $^1$ Junior fellow of the research council (DOC)
of the K.U. Leuven.\\
\hspace*{5pt} E-mail: Friedemann.Brandt@fys.kuleuven.ac.be\\
\noindent $^2$ E-mail: henneaux@ulb.ac.be\\
\noindent $^3$ E-mail: awilch@ulb.ac.be}

\end{titlepage}

\section{Introduction}

This letter is devoted to the implementation of global ($\equiv$ rigid)
symmetries in the antifield-formalism.

Consider a gauge-invariant theory with gauge-invariant local action
\beqn
   I  =  \vol \, {\cal L}
   \left([\phi],x\right)
\label{r1}
\eeqn
and gauge-symmetries
\beqn
\delta_{\eta} \;\phi^i=
\sum_{k=0}^t r^{i\;\mu_1 \cdots \mu_k}_{\alpha}\left([\phi],x\right)
\p_{\mu_1 \cdots \mu_k}\eta^{\alpha}
\label{r2}
\eeqn
where $ \eta^{\alpha} $ are arbitrary space-time functions
and where the argument $[\phi]$ indicates dependence
on the fields and on their derivatives $\p_\mu\phi^i$, \ldots,
$\p_{\mu_1\cdots\mu_r}\phi^i$ up to some finite (but arbitrary)
order $r<\infty$.
It has been established in \cite{a1} that
under quite general regularity
conditions on the Lagrangian, the
gauge-symmetries and the reducibility
functions (if any), there exists a
local solution of the classical master
equation introduced by Zinn-Justin
as well as Batalin and Vilkovisky 
\cite{a2,a2bis,a3},
\beqn
\left( S_0, S_0 \right)  =  0,\quad
S_0  =  I \; + \; \mbox{ ``more'' }\;,
\label{r3}
\eeqn
where  ``more''  is the integral of a
local function that involves at least
one ghost and one antifield in each of its field monomials.
The proof given in \cite{a1} follows the lines
of homological perturbation theory
\cite{a4} and adapts to local functionals
the proofs given in \cite{a5,a6,a7}
for irreducible gauge-theories and in
\cite{a8} for reducible ones (for a
general discussion of the homological
tools underlying the antifield formalism, see \cite{a9}).

Besides the gauge-symmetries (\ref{r2}),
the theory may possess also global
symmetries (e.g. Poincar\'{e}-symmetry,
rigid supersymmetry, global internal
symmetries etc.),
\beqn
\delta_{\GP}\;\phi^i  =  t_A^i\left([\phi],x\right) \GP^A
\equiv \left( \delta_A \phi^i \right)  \GP^A \; ,
\label{r5}
\eeqn
where $ \GP^A $ are constant parameters.
The action (\ref{r1}) is invariant
under (\ref{r5}),
\beqn
   \delta_{\GP} \;I \; = \; 0 \; .
\label{r6}
\eeqn
The rigid symmetry (\ref{r5}) may or may
not be linearly realized, i.e. the 
functions $ t^i_A $ need not be linear
(homogeneous) in the fields and their
derivatives. Furthermore, the commutator
of a rigid symmetry with a
gauge-symmetry may involve both a non-zero
gauge-symmetry and
on-shell trivial
gauge-symmetries, while the rigid symmetries
may close only up to a
gauge-symmetry and on-shell trivial
transformations. Schematically,
\barr  
    \left[ \delta_A, \delta_{\alpha} \right] \; \phi^i & = &
    \mu^{\beta}_{A \alpha} \; R^i_{\beta} \; + \; 
    \mu^{ij}_{A \alpha} \; \frac{\delta I}{\delta \phi^j}\ ,
\label{r7}
\\[3ex]
    \left[ \delta_A, \delta_B \right] \; \phi^i & = &
    \f BAC \; \delta_C \; \phi^i \; + \; 
    \lambda_{BA}^{\alpha} \; R^i_{\alpha} \; + \; 
    \lambda^{ij}_{BA} \; \frac{\delta I}{\delta \phi^j} \; ,
\label{r8}
\earr
where we have set 
$ \delta_{\eta} \, \phi^i \equiv \eta^{\alpha} R^i_{\alpha}
\equiv \eta^\alpha\delta_\alpha\phi^i$
and where we have used temporarily DeWitt's condensed notations.
In (\ref{r8}), the $ \f ABC $ are the
{\it structure constants} of the (graded) rigid
symmetry  Lie algebra.

The first point investigated in this letter is
whether the solution $ S_0 $
of the master equation (\ref{r3}) remains
invariant under the global symmetry
(\ref{r5}). More precisely: 
Is it possible to modify the transformation law (\ref{r5})
by ghost dependent contributions
(so that (\ref{r5}) is unchanged if the ghosts are set
equal to zero) and define the transformation
rules for the ghosts and the
antifields in such a way that $ \delta_{\GP}\,S_0  =  0 $ ?
The answer to this question turns out to be
always affirmative and,
furthermore, the global symmetry is
canonically generated in the antibracked
by a {\it local} generator $ S_A $ of ghost-number minus one
\beqn
     S_A \; = \; \vol \left( \phi_i^{*}\;t^i_A \; + \; 
     \mbox{``more''} \right)
\label{r9}
\eeqn
such that
\beqn
    \delta_{\GP}\;z^{\Delta} \; = \; \left(z^{\Delta}, S_A \right)
    \GP^A
\label{r10}
\eeqn
for all variables $ z^{\Delta} $,
including the ghosts and the antifields,
and
\beqn
    \left( S_0, S_A \right)\GP^A \; = \;
    \delta_{\GP}\;S_0 \; = \; 0 \; .
\label{r11}
\eeqn

In order to analyse whether a global
symmetry is quantum-mechanically
anomalous, and if not, how it gets
renormalized, it is convenient to 
consider an extended generating
functional $ S $ that incorporates both
the gauge-symmetries and the global symmetry,
\beqn
    S \; = \; S_0 \; + \;S_A\;\xi^A\; +
    \s0 12\;S_{AB}\;\xi^B\;\xi^A \; + \;
    O \left( \xi^3 \right) \; ,
\label{r12}
\eeqn
where $ \xi^A $ are constant ghosts
associated with the global symmetry,
such that 
\beqn
    \left( S, S \right) \; + \;\frac{\p^R S}{\p \xi^C}\;
    \f BAC\; \xi^A\;\xi^B \;\left(-1\right)^{\e_B} \; = \; 0
\label{r13}
\eeqn
(``extended master equation''). Here, $ \e_A $ is the
parity of the global symmetry $\delta_A$
and the $ \f ABC $ are the structure
constants appearing in (\ref{r8}).
Equation (\ref{r13}) guarantees the existence of
a nilpotent antiderivation $D$ ($D^2=0$) that encodes both the
gauge and global symmetries and is defined on
any function(al) $X$ of the fields, antifields and
constant ghosts by
\beqn D \, X\equiv \left( X, S \right) +\s0 12
  \left(-1\right)^{\e_B} \frac{\p^RX}{\p\xi^C}\;
  \f BAC\;\xi^A\;\xi^B\; .
\label{extbrs}
\eeqn
Such an extended generator $ S $ or antiderivation $D$ have been
constructed by various authors mostly (but not only) in the context of
globally supersymmetric theories \cite{a2bis,a15_1,various}.
It has also been analysed recently in \cite{a19} in connection with 
equivariant cohomology.
 
Now, the second question addressed in
this letter is whether the existence
of a {\it local} solution of the extended
master equation (\ref{r13}) starting
like (\ref{r12}) is always guaranteed.
It is shown that the answer to this
question may be negative whenever the
theory has higher order non-trivial
conservation laws,
\beqn
\p_{\mu_1} j^{\, \left[\mu_1 \cdots \mu_k\right]}
\left([\phi],x\right) \approx 0 \; ,\quad
j^{\,\left[\mu_1 \cdots \mu_k\right]}\left([\phi],x\right)
\not\approx
\p_{\mu_0} \omega^{\,\left[\mu_0 \cdots \mu_k\right]}
\left([\phi],x\right)
\quad (k \geq 2) ,
\label{r14} 
\eeqn
where $\approx$ denotes weak ($\equiv$ on-shell) equality and
$\left[\mu_1 \cdots \mu_k\right]$ complete antisymmetrization.
Thus, even though any global symmetry can be extended to the space of
the ghosts and antifields in such a way that it leaves the solution
$ S_0 $ of the ``restricted master equation'' (\ref{r3}) invariant, it
may be impossible to complete $ S_0+ S_A\xi^A $ to a local solution
$S$ of equation (\ref{r13}). We provide explicit examples where
obstructions arise. We also
give conditions for the expansion (\ref{r12})
to be unobstructed so that 
the extended formalism based on (\ref{r13}) exists.
These conditions are met,
for example, in Yang-Mills gauge-theories, gravity as well as 
Super-Yang-Mills models. [As usual in the
context of the antifield formalism,
we say that a functional is local if each
term in its expansion according
to the antighost-number is the integral of a function of the fields, the 
ghosts, the antifields and a finite number of their derivatives
(``local function'').]

\section{Implementation of Global Symmetries in the Antifield Formalism}

Our first task is to investigate whether the global symmetry (\ref{r5})
of the classical action $ I $ is also a symmetry of the solution $ S_0 $
of the master equation (\ref{r3}). Since $ S_0 $ involves more variables
than $ I $ does, what we really mean by this question is whether one can
modify the global symmetry (\ref{r5}) by ghost
dependent terms (invisible when the
ghosts are set equal to zero) and
define appropriate transformation rules
for the ghosts and the antifields in such a way that 
$ \delta_{\GP}  S_0 = 0 $.
It cannot be stressed  enough that any ``proof'' of invariance of $ S_0$
under a given global symmetry that does not
indicate at the same time how
to transform the ghosts and antifields is incomplete and meaningless.

We claim that the answer to this first question is always positive and
that, moreover, the symmetry in the extended
space is canonically generated
in the antibracket, as in equation (\ref{r10}).
To prove this statement, let us construct
directly the canonical generator
$ S_A $.

In order to reproduce (\ref{r5}) through (\ref{r10}),
it is necessary that 
$ S_A $ starts like in equation (\ref{r9}),
where ``more'' contains at least one local
ghost of the gauge-symmetry
and  accordingly has antighost-number greater than one.
(For more information
on our grading conventions and on the Koszul-Tate differential in the
antifield formalism used below, see \cite{a8,a9}.)
Now, the equation $ \delta_{\GP}\;I = 0 $
is equivalent, in terms of the
Koszul-Tate differential $ \delta $, to the condition
$ \vol \;\delta \left( \phi_i^{*}\;t^i_A \right) = 0 $, i.e., 
$ \delta \left( \phi_i^{*}\;t^i_A \right) \; + \; \p_{\mu} j^{\mu} = 0 $
for some $ j^{\mu} $.
This means that the first term $ \phi_i^{*}\;t^i_A $ of
$ S_A $ defines a
cocycle of the cohomology $ H_1\left(\delta|d\right) $,
which is non-trivial
because we assume the given rigid symmetry to be itself non-trivial.
By using Theorem 6.1 of reference \cite{a20} on the
isomorphism between
$ H_k\left(\delta|d\right) $ and the BRST-cohomology
modulo $ d $ at negative
ghost number $ - k $,
\beqn
   H^{-k}\left(s|d\right) \; \simeq \; H_k\left(\delta|d\right) 
   \quad (k > 0) \; ,
\label{s1}
\eeqn
it is possible to infer the existence of a local
``BRST-invariant extension''
$ S_A $ of $ \vol \left( \phi_i^{*}\;t^i_A \right) $ with the required
property
\beqn
    s\;S_A \; = \; 0 \; .
\label{s2}
\eeqn
Here, $ s $ is the BRST differential associated with the
{\it gauge}-symmetry, $ sA = \left(A, S_0\right) $.
The proof of (\ref{s1}) and of
the resulting existence of $ S_A $ uses the
acyclicity of $ \delta $ in the
space of local functionals containing at least one local ghost of the
gauge-symmetry and one antifield in each term \cite{a1}.

The equation (\ref{s2}) can be rewritten as 
\beqn
    \left(S_A, S_0\right) \; = \; 0
\label{s3}
\eeqn
since the BRST-transformation $ s $ is
canonically generated by $ S_0 $.
If one defines the transformation rules for
all the variables according to
(\ref{r10}), one sees that equation (\ref{s3}),
which expresses the BRST-invariance
of $ S_A $, can be read backwards and expresses also the invariance of 
$ S_0 $ under the symmetry generated by $ S_A $. This answers positively
the first question raised above. Note that since
$ S_A $ is a local functional,
the transformation rules of all the variables $ z^{\Delta} $,
$ \delta_{\GP}z^{\Delta} = \left(z^{\Delta}, S_A \right)\GP^A $,
are local functions.

It is well known that the solution $S_0$
of the master equation (\ref{r3}) carries
some ambiguity. In the recursive construction of $ S_0 $, one has the
possibility to add at each stage an arbitrary $ \delta $-exact term.
One could choose these higher order terms involving the ghosts and the
antifields in a manner that could conflict with some preconceived idea
of manifest invariance under the global symmetry of the theory.
What our result indicates is that this does not matter.
One can choose the higher order terms in a way which is not manifestly
invariant since it is always possible to extend the transformation laws
in the space of the fields, the ghosts and the antifields so
that $ S_0 $ is strictly invariant, no matter how one has fixed
the ambiguity in $ S_0 $.

By using (\ref{r7}), one verifies easily that in $S_A$,
\begin{eqnarray*}
S_A  &=&  S_{A, (1)}  + S_{A, (2)} \; + \cdots\ ,\\
S_{A, (1)} & =& \vol \;\phi_i^{*}\;t^i_A\ ,
\end{eqnarray*}
the term of antighost-number two is given by
\beqn
    S_{A, (2)} \, \sim\,
          C^{*}_{\beta}\;\mu_{A \alpha}^{\beta} \;C^{\alpha} \; + \;
          \s0 12\;\phi^{*}_i\;\phi^{*}_j\;\mu_{A \alpha}^{ij} \;
          C^{\alpha} \; ,
\label{s6}
\eeqn
where we use again the condensed notation (and `$\sim$' to indicate
that parity dependent phase factors are suppressed).
The first term on the right hand side of (\ref{s6}) determines the 
transformation rule for the local ghosts to first order;
the second term
modifies the transformation rule of the fields $ \phi^i $ by the term
$ \phi^{*}_j\;\mu_{A \alpha}^{ij} \;C^{\alpha} $ and,
according to (\ref{r7}),
arises only if the commutator of a rigid symmetry with a gauge-symmetry
involves an on-shell trivial symmetry.

Since the $ S_A $ are BRST-closed,
their antibracket is also BRST-closed.
By using (\ref{r8}), one finds that 
$ \left(S_A, S_B\right) -  \f ABC\;S_C  $
is a BRST cocycle of ghost number minus
one whose component of antighost-number one is $ \delta $-exact 
in the space of local functionals,
\barr
   \left(S_A, S_B\right)- \f ABC\;S_C
   & \sim &
   \delta\; 
   \left(
         C^{*}_{\alpha}\;\lambda^{\alpha}_{AB} + 
         \s0 12\;\phi^{*}_i\;\phi^{*}_j\;\lambda_{AB}^{ij}
   \right)  \nn
\\
   & & \mbox{\hspace{0.5cm}} +
   \mbox{terms of higher antighost-number.} \nn
\earr
Thus, according to the results of \cite{a20} and \cite{a1},
$ \left(S_A, S_B\right) -\f ABC S_C $
is actually $ s $-exact,
i.e. there exists a local functional $ S_{AB} $ of ghost-number $-2$
such that
\beqn
    \left(S_A, S_B\right)  - 
    \f ABC S_C  -  (-1)^{\e_A}
    \left(S_0, S_{AB}\right)  =  0.
\label{s7}
\eeqn
If one redefines $ S_A $ as 
$ S_A \longrightarrow S_A  +  \left(S_0, K_A\right) $, where $ K_A $
has ghost-number $ -2 $, which is tantamount to redefining the
global symmetry by gauge-symmetries and on-shell trivial symmetries,
one finds that $ S_{AB} $ transforms as  
\[ 
  S_{AB} \longrightarrow S_{AB}
  -(-1)^{\e_A}(\f ABC K_C+M_{AB}-(-1)^{\e_A\e_B}M_{BA})
\]
with
$ M_{AB}=(K_A,S_B)-\s0 12(K_A,(S_0,K_B))$.

The above derivation mimics exactly the treatment of \cite{a21} of rigid
symmetries in the Hamiltonian version of BRST theory, where the term
$ \left(S_{AB}, S_0\right) $ was called the ``BRST extension'' of the
Lie algebra defined by $ \f ABC $. In fact, it is possible to
apply the antifield
approach also to the Hamiltonian formalism \cite{a22,a23,a24}. 
Doing this
in the present context, one finds for $ S_A $ a formula analogous to the
formula connecting $ S_0 $ and the canonical BRST generator
$ \Omega $ \cite{a22,a23}, i.e.
\beqn
   S_A  \sim  \int \mbox{dt} \Bigg( 
   q^{*}_i\left[q^i, Q_A\right]  +  p^{*i}\left[p_i, Q_A\right]
   + \lambda^a\left[\varrho_a, Q_A\right] + 
   \eta^{*}_a\left[\eta^a, Q_A\right]\Bigg)  ,
\label{s8}
\eeqn
where $ Q_A $ are the canonical generators of the rigid symmetry in the
extended phase space.

\section{Extended Master Equation}

We now turn to the question whether a
local solution to the master equation
(\ref{r13}) incorporating the global symmetries is guaranteed to exist.
As we shall show, the answer may be negative.

To analyse the question, we expand the
searched-for $ S $ in powers of the 
global ghosts $ \xi^A $ as in \cite{a19} (see equation (\ref{r12})).
The coefficients $ S_A $, $ S_{AB} $, $ S_{ABC} \; \ldots $
should be local
functionals of decreasing ghost-number $ -1,\; -2,\; -3,\; \ldots $.
With $ S_A $ and $ S_{AB} $ constructed as above, the extended master 
equation holds up to order $ \xi^2 $ included, i.e.
\[
\left(S^{(2)},S^{(2)}\right)+
\frac{\p^R S^{(2)}}{\p \xi^C}\;
    \f BAC\; \xi^A\;\xi^B \;\left(-1\right)^{\e_B}
    =O \left( \xi^3 \right)
\]
where $S^{(2)}=S_0  +  S_A\xi^A  + \s0 12 S_{AB}\xi^B\xi^A$.
Therefore, let us proceed recursively. Assume that one has constructed
$ S $ up to order $ \xi^k $, 
$ S^{(k)} = S_0 + S_A\;\xi^A + \cdots +
\frac {1}{k!}S_{A_1\cdots A_k} \;
  \xi^{A_k}\ldots\xi^{A_1} $,
so that the extended master equation  holds up to
order $ k $ included, and let us try to
determine the term of order $ k + 1 $
in $ S $ so that $ S^{(k+1)}$
solves the extended master equation (\ref{r13}) up to order $ k + 1 $
included. 

The equation for $ S_{A_1\cdots A_{k+1}} $ that follows from (\ref{r13})
takes the form 
\beqn
   \left( S_{A_1\cdots A_{k+1}}, S_0 \right) \; = \;
   R_{A_1\cdots A_{k+1}} \; ,
\label{m2}
\eeqn
where the local functional $ R_{A_1\cdots A_{k+1}} $ has ghost-number
$ -k $ and is built out of the already constructed $ S_{A_1\cdots A_j}$.
Thus, in order for $ S_{A_1\cdots A_{k+1}} $ to exist,
$ R_{A_1\cdots A_{k+1}} $ should be $ s $-exact in the space of local
functionals.
By using the Jacobi-identity for the antibracket, one easily checks that
$ R_{A_1\cdots A_{k+1}} $ is $ s $-closed.
The proof follows the standard pattern of
homological perturbation theory
\cite{a4,a9} and will not be repeated here. Accordingly, a sufficient
condition for 
$ S_{ABC},\; S_{ABCD}, \ldots , S_{A_1\cdots A_{k+1}}, \ldots $
to exist is that the BRST cohomological groups in the space of local
functionals, denoted by $ H^j\left(s|d\right) $, vanish%
\footnote{Strictly speaking, the cohomology
  $H^j\left(s\right)$ of $s$ in
  the space of local functionals is not exactly the same
  as the cohomology
  $H^j\left(s|d\right)$ in the space of local volume forms
  because the surface
  terms that one drops in calculating
  $H^j\left(s|d\right)$ may be non zero.
  Thus, a cocycle of $H^j\left(s|d\right)$ does not yield necessarily a
  cocycle of $H^j\left(s\right)$ upon integration.
  However, when discussing
  locality, it is really $H^j\left(s|d\right)$ that is relevant.
  Hence, we
  deal exclusively from now on with $H^j\left(s|d\right)$.}
for $ j \leq -2 $.

Now, as we have already recalled, one has 
$ H^{-k}\left(s|d\right) \simeq H_k\left(\delta|d\right) $, where 
$ \delta $ is the Koszul-Tate differential
associated with the equations
of motion. Furthermore, $ H_k\left(\delta|d\right) $
is itself isomorphic
to the characteristic cohomology
$ H_{char}^{n - k}\left(d\right) $ of
Vinogradov \cite{a25} and
Bryant and Griffiths \cite{a26}, which describes
the higher-order conservation laws 
(\ref{r14}) \cite{a20}.
Thus, obstructions to the existence of the higher-order terms
$ S_{A_1\cdots A_{k + 1}} $,
i.e. to the existence of a local functional
solution of the extended master equation (\ref{r13}),
may arise whenever there
exist non-trivial higher-order conservation laws in the theory.

Of course, the presence of higher-order conservation laws does not
necessarily lead to obstructions.
It could happen that $ R_{A_1\cdots A_{k+1}} $ in (\ref{m2})
is always in the trivial class of $ H^{-k}\left(s|d\right) $.
This would be for instance the
case if $ R_{A_1\cdots A_{k+1}} $ contained
no term of antighost-number $ k $.
Our main result, however,
is that there exist global symmetries for which
the obstructions are effectively present%
\footnote{Note that the term of antighost-number $ k $ of
   $ R_{A_1\cdots A_{k+1}} $
   does not involve the local ghosts associated
   with the gauge-symmetries,
   so the argument of \cite{a1} on the vanishing
   of $ H_i\left(\delta|d\right)$ ($i \geq 1) $ when local ghosts are
   present does not apply.}.
We shall establish this point by means of an explicit example.

A theory with a non-vanishing second homology group 
$ H_2\left(\delta|d\right) $ is the free Maxwell-theory without sources.
In that case, $ H_2\left(\delta|d\right) $ is one-dimensional
(except in two dimensions). One may
take as representative of the non-trivial cohomology class of
$ H_2\left(\delta|d\right) $ the antifield $ C^{*} $ associated
with the ghost field $ C $ ($\delta C^{*} = -\p_{\mu}A^{*\mu}) $.
This class corresponds to the conservation law
$ \p_{\mu}F^{\mu\nu} \approx 0 $.

The Maxwell action in Minkowski-space is invariant,
among other symmetries,
under translations,
\beqn
   \delta_a \; A_{\mu} \; = \; a^{\nu}\p_{\nu} A_{\mu} \; ,
\label{m3}
\eeqn
and under the following $x$-dependent shifts in the fields,
\beqn
   \delta_{\platz}\;A_{\mu} \; = \; \platz_{\mu \nu}\;x^{\nu}\;,
   \mbox{\hspace{2cm}}
   \platz_{\mu \nu} = - \platz_{\nu \mu}
\label{m4}
\eeqn
where $a^\mu$ and $b_{\mu\nu}$ are constant parameters.
[The symmetric part of $ \platz_{\mu \nu} $ defines a gauge-symmetry
and is excluded from the discussion for that reason.] The symmetries
(\ref{m3}) and (\ref{m4}) commute up to a gauge-transformation.
Thus, $ \f ABC = 0 $. We shall show that the construction of a
solution of the extended master equation (\ref{r13})
incorporating (\ref{m3}) and 
(\ref{m4}) is obstructed.

The solution of the ``restricted'' master equation (\ref{r3}) reads
\beqn
   S_0 = \vol \left(
                    -\s0 14\;F_{\mu \nu}\;F^{\mu \nu} +
                    A^{*\mu}\;\p_{\mu}C
              \right) \;, 
   \mbox{\hspace{1cm}}
   F_{\mu \nu} = \p_{\mu} A_{\nu} - \p_{\nu} A_{\mu} \; .
\label{m5}
\eeqn
The BRST-invariant term 
$ S_1 \equiv \xi^A S_A$ generating the global symmetries (\ref{m3}) and
(\ref{m4}) is
\beqn
   S_1 = \vol \left[ A^{*\mu}
         \left(\xi^{\nu}\,\p_{\nu} A_{\mu} +
         \xi_{\mu \nu}\, x^{\nu} \right)
         \; - \; C^{*}\;\xi^{\mu}\p_{\mu}C \right] \; ,
\label{m6}
\eeqn
where $ \xi^{\nu} $ and $ \xi_{\mu \nu} $ are constant anticommuting
ghosts associated with the translations (\ref{m3}) and with the symmetry
(\ref{m4}) respectively. The generator (\ref{m6})
determines the transformation properties
of the antifields and the ghosts under the given rigid symmetries.
By computing their antibracket with $ S_1 $,
one finds that these do not
transform under (\ref{m4}), while they behave as
\[
\delta_a z^\Delta=a^\mu \p_\mu z^\Delta\quad\quad
\forall\; z^\Delta\in
\{C,\; A^{*\mu},\; C^*\}
\]
under translations, as expected.

Although one could add to $ S_1 $ a BRST-exact functional, we shall not
do it here. This is because we want to stick to the original form 
(\ref{m3}) and (\ref{m4}) of the rigid symmetries,
without modifying them by
the addition of gauge-transformations or on-shell trivial symmetries.
Also, we want to maintain the quadratic character of $ S_1 $.

With $ \f ABC = 0 $, the extended master equation reads simply
$ \left( S,S \right) = 0 $.
We are looking for a solution of that equation
of the form $ S = S_0 + S_1 + S_2 + \cdots $,
where $ S_k $ has degree $k$
in the constant ghosts $ \xi^{\nu} $ and $ \xi_{\mu \nu} $.
The extended master equation requires
$ \left(S_1,S_1\right) + 2\;\left(S_0,S_2\right) = 0 $ for some $ S_2 $.
This implies
\beqn
   S_2 = \vol \left(C^{*}\; x^{\mu}\;\xi_{\mu \nu}\;\xi^{\nu} \right)
\label{m7}
\eeqn
up to a BRST-closed term, which reflects the fact that the commutator of
the global symmetries (\ref{m3}) and
(\ref{m4}) is a gauge-transformation with parameter
$ a^\mu\platz_{\mu \nu}x^{\nu} $.

It is for the next term $ S_3 $ that one meets the obstruction:
one finds indeed
\beqn
   \left(S_1,S_2\right)  = \xi^{\mu}\;\xi^{\nu}\;\xi_{\mu \nu}
   \vol\, C^{*}
\label{m8}
\eeqn 
the integrand of which is non-trivial in $ H_2\left(\delta|d\right) $.
Therefore, there is no $ S_3 $ such that 
$ \left(S_1,S_2\right) + \left(S_0,S_3\right) = 0 $ holds, i.e. a local
solution of the master equation (\ref{r13})
encoding the above rigid symmetries
{\it simply does not exist}. Note that the ambiguity in $ S_2 $
[$S_2 \longrightarrow S_2 + M_2 $ where $ M_2 $ is BRST-closed
and of antighost-number at least two, i.e.
$S_2 \longrightarrow S_2 + f(\xi) \vol\, C^{*}
+ \left(S_0,K_2\right) $]
does not allow for the removal of the obstruction (\ref{m8}).

We note however that one can indeed remove the obstruction by
further extending the formalism and the master equation
(\ref{r13}).
To that end we introduce another constant ghost,
$Q$, associated with the global
reducibility identity on the gauge symmetry responsible for
the non-vanishing of $H_2(s|d)$ \cite{a20}. We assign to it ghost
number 2 and even Grassmann parity.
We also introduce a constant anti`field' $Q^*$ conjugate
to $Q$.
Then, with $S_0$, $S_1$ and $S_2$ as above,
\[
S=S_0+S_1+S_2+\vol\, C^*Q-Q^*\,
\xi^{\mu}\xi^{\nu}\xi_{\mu \nu}
\]
solves an extended master equation in the standard form,
$(S,S)=0$, where the antibracket now involves also
(ordinary) derivatives w.r.t.\ $Q$ and $Q^*$
(analogously one can always cast (\ref{r13}) in the
standard form by introducing $\xi^*_A$).

The example can be generalized to free 2-form gauge-fields, for which 
$ H_2\left(\delta|d\right) = 0 $
but $ H_3\left(\delta|d\right) \not= 0 $
\cite{a27}. The symmetries 
$ B_{\mu \nu} \longrightarrow  B_{\mu \nu} + 
  a^{\varrho}\p_{\varrho}B_{\mu\nu} +
  \platz_{\mu \nu \varrho}x^{\varrho} $
(with $ \platz_{\mu \nu \varrho} $ completely antisymmetric)
can be incorporated
in the master equation (\ref{r13}) up to order $ S_3 $ included
($ H_2\left(\delta|d\right) = 0 $) but get obstructed at the next level.
Again, one can remove the obstruction by introducing a constant ghost
with ghost number 3 and odd Grassman parity.
One finds similar results for free $p$-form gauge-fields with $p>2$.

\section{Conclusions}

In this letter, we have proved that any
global symmetry of a gauge-invariant
theory can always be extended to the antifields
and the ghosts so as to be a
symmetry of the restricted (usual) master equation (\ref{r3}).
We emphasize that this holds for the solution
of the master equation
{\em before} gauge fixing and that, in general, our result does not
imply an analogous property for the gauge fixed action
(see \cite{a15_1} for a discussion of this problem).
We have then discussed
the incorporation of the rigid symmetries
in the extended formalism and have
shown that one cannot take for granted the
existence of a local solution
of the extended master equation (\ref{r13})
in which the rigid symmetries
are included with constant ghosts. Indeed,
one can meet obstructions and
we have provided explicit examples for that.

The possible obstructions are given by the cohomological groups
$ H_i\left(\delta|d\right) $, $ \left( i \geq 2 \right) $ of the
characteristic cohomology. Thus, when
$ H_i\left(\delta|d\right) = 0 $ for all $i \geq 2 $, the
global symmetries can all be incorporated in the master equation
(\ref{r13}).
This is the case for the most interesting physical theories, since, for
example,
$ H_i\left(\delta|d\right) = 0 $ $ \forall i \geq 2  $ for
Yang-Mills theories with a semi-simple gauge-group, their supersymmetric
extensions and also for gravity \cite{a20}.
We stress, however, that even in those cases, the existence
of the extended formalism based on
equation (\ref{r13}) is a nontrivial property which is
not automatic and needed demonstration as the above counterexamples
indicate.

Finally, we have shown how one can remove the obstructions
in all these counterexamples by further extending the
formalism. 
In fact this just illustrates the general case: one can
set up an extended antifield formalism, generalizing
the one considered here, where obstructions are absent
and the higher order conservation laws are encoded in the
formalism too \cite{a28}.

Acknowledgements:
This work has been supported  in part by research
funds from the F.N.R.S.
(Belgium) and a research contract with the Commission of the European 
Community.


\begin{thebibliography}{60}

\bibitem{a1} M. Henneaux, Commun. Math. Phys. 140 (1991) 1.

\bibitem{a2} J. Zinn-Justin, ``Renormalisation of Gauge Symmetries'',
             in:
             Trends in Elementary Particle Theory, 
             Lecture Notes in Physics 37, Springer,
             Berlin, Heidelberg, New York 1975.

\bibitem{a2bis} J. Zinn-Justin,
             ``Quantum Field Theory and Critical Phenomena'',
             2nd Edition,
             Claradon Press, Oxford 1993.

\bibitem{a3} I. A. Batalin and G. A. Vilkovisky, 
             Phys. Lett. B 102 (1981) 27, Phys. Rev. D 28 (1983) 2567,
             Phys. Rev. D 30 (1984) 508.

\bibitem{a4} G. Hirsch, Bull. Soc. Math. Belg. 6 (1953) 79;\\
J.D. Stasheff, Trans. Am. Math. Soc. 108 (1963) 215,293;\\
V.K.A.M. Gugenheim, J. Pure Appl. Alg. 25 (1982) 197;\\
V.K.A.M. Gugenheim and J.D. Stasheff, Bull. Soc. Math. Belg.
38 (1986) 237.

\bibitem{a5} B. de Wit and J. W. van Holten,
             Phys. Lett. B 79 (1978) 389.

\bibitem{a6} B. L. Voronov and I. V. Tyutin,
             Theor. Math. Phys. 50 (1982) 218.

\bibitem{a7} I. A. Batalin and G. A. Vilkovisky,
             J. Math. Phys. 26 (1985) 172.

\bibitem{a8} J. M. L. Fisch and M. Henneaux, 
             Commun. Math. Phys. 128 (1990) 627;
             This paper is itself based on
             J. Fisch, M. Henneaux, J. Stasheff
             and C. Teitelboim, Commun. Math. Phys. 120 (1989) 379.

\bibitem{a9} M. Henneaux, Nucl. Phys. B (Proc. Suppl.) 18A (1990) 47;\\
             M. Henneaux and C. Teitelboim,
             ``Quantization of Gauge Systems'',
             Princeton University Press, Princeton (1992).

\bibitem{a10} P. Breitenlohner and D. Maison, ``Renormalisation of 
              Supersymmetric Yang-Mills Theories'', 
              in: Cambridge 1985, Proceedings: ``Supersymmetry and its
              Applications'', p.309.

\bibitem{a15_1} J. W. van Holten, Phys. Lett. B 200 (1988) 507.

\bibitem{various}
L. Bonora, P. Pasti and M. Tonin, Phys. Lett. B 156 (1985) 341;
\\
C. Becchi, A. Blasi, G. Bonneau, R. Collina and F. Delduc,
              Commun. Math. Phys. 120 (1988) 121;
\\
R. Kaiser, Z. Phys. C 39 (1988) 585;
\\
P. Altevogt and R. Kaiser, Z. Phys. C 43 (1989) 455;
\\
P. Howe, U. Lindstrom and P. White, Phys. Lett. B 246 (1990) 430;
\\
L. Baulieu, M. Bellon, S. Ouvry and J.-C. Wallet,
              Phys. Lett. B 252 (1990) 387;
\\
J. A. Dixon, Commun. Math. Phys. 140 (1991) 169;
\\
P. L. White, Class. Quant. Grav. 9 (1992) 413, 1663;
\\
F. Brandt, Nucl. Phys. B 392 (1993) 428; Phys. Lett. B 320 (1994) 57;
\\
G. Bonneau, Phys. Lett. B 333 (1994) 46;
              Helv. Phys. Acta 67 (1994) 930, 954;
\\
N. Maggiore, O. Piguet and S. Wolf, Nucl. Phys. B 458 (1994) 403;
\\
N. Maggiore, Int. J. Mod. Phys. A 10 (1995) 3781, 3937;
\\
N. Maggiore, O. Piguet and M. Ribordy, Helv. Phys. Acta 68 (1995) 264;
\\
O. Piguet and S. P. Sorella, ``Algebraic Renormalisation'',
   Lecture Notes in Physics, Vol. m28, Springer Verlag, Berlin,
   Heidelberg, 1995.

\bibitem{a19} F. Delduc, N. Maggiore, O. Piguet and S. Wolf,
              hep-th/9605158.

\bibitem{a20} G. Barnich, F. Brandt and M. Henneaux, 
              Commun. Math. Phys. 174 (1995) 57.

\bibitem{a21} M. Henneaux, Nucl. Phys. B 308 (1988) 619;
              and contribution in ``Geometrical and Algebraical Aspects
              of Nonlinear Field Theory'', p. 119,
              S. De Filippo, M. Marinoro, G. Marmo and G. Vilasi eds,
              Elsevier Science Publishers BV
              (North Holland, Amsterdam (1989)).

\bibitem{a22} J. M. L. Fisch and M. Henneaux,
              Phys. Lett. B 226 (1989) 80.

\bibitem{a23} W. Siegel, Int. J. Mod. Phys. A 4 (1989) 3951.

\bibitem{a24} C. Battle, J. Gomis, J. Paris and J. Roca, 
              Phys. Lett. B 224 (1989) 288;
              Nucl. Phys. B 329 (1990) 139.

\bibitem{a25} A. M. Vinogradov, Sov. Math. Dokl. 18 (1977) 1200,
              19 (1978) 144, 19 (1978) 1220.

\bibitem{a26} R. L. Bryant and P. A. Griffiths, ``Characteristic
              Cohomology of Differential Systems (I): General Theory'',
              Duke University Mathematics Preprint Series,
              volume 1993 $n^0$ 1,
              January 1993.

\bibitem{a27} M. Henneaux, B. Knaepen and C. Schomblond,
              in preparation.

\bibitem{a28} F. Brandt, M. Henneaux and A. Wilch,
              in preparation.

\end{thebibliography}
\end{document}